\documentclass[showpacs,twocolumn,showkeys,amsmath,amssymb,prx]{revtex4-1}

\usepackage{bm}
\usepackage{bbm}
\usepackage{dsfont}
\usepackage{color}
\usepackage{graphicx}   
\newcommand{\rd}{{\mathrm d}}
\newcommand{\re}{{\mathrm e}}
\newcommand{\ri}{{\mathrm i}} 
\newcommand{\lF}{\langle\!\langle}
\newcommand{\rF}{\rangle\!\rangle}
\begin{document}

\title[Following Floquet states]
      {Following Floquet states in high-dimensional Hilbert spaces}  

\author{Nils Kr\"uger and Martin Holthaus}
	
\affiliation{Carl von Ossietzky Universit\"at, Institut f\"ur Physik,
	D-26111 Oldenburg, Germany}	
                   
\date{November 07, 2021}

\begin{abstract}
An iterative algorithm is established which enables one to compute individual 
Floquet states even for many-body systems with high-dimensional Hilbert spaces 
that are not accessible to commonly employed conventional methods. A strategy 
is proposed for following a Floquet state in response to small changes of a 
given system's Hamiltonian. The scheme is applied to a periodically driven 
Bose-Hubbard chain, verifying the possibility of pseudoadiabatic Floquet state
following. In particular, it is demonstrated that a driving-induced Mott 
insulatorlike target Floquet state can be populated with high efficiency if 
the driving amplitude is turned on smoothly but not too slowly. We conclude 
that the algorithm constitutes a powerful tool for the future investigation 
of many-body Floquet systems.         
\end{abstract} 

\keywords{Periodically driven quantum systems, Floquet states, 
	largest eigenvalue, Bose-Hubbard model}

\maketitle 

%%%%%%%%%%%%%%%%%%%%%%%%%%%%%%%%%%%%%%%%%%%%%%%%%%%%%%%%%%%%%%%%%%%%%%%%%%%%%%%%

\section{Introduction}
\label{S_1}

The experimental and theoretical investigation of periodically time-dependent 
many-body quantum systems, nowadays often dubbed Floquet systems, has turned 
into a remarkably fruitful area of physics in recent years, comprising, among 
many others, the dynamics of cold atomic quantum gases in periodically driven 
optical lattices~\cite{Eckardt17}, the principles underlying Floquet time 
crystals~\cite{ZhangEtAl17,ChoiEtAl17}, and fundamental aspects of 
nonequilibrium statistical physics~\cite{RubioAbadalEtAl20,LuitzEtAl20}.    

Advances on the theoretical side of this field appear to be impeded by 
the fact that, while it is still feasible to solve the time-dependent 
Schr\"odinger equation numerically for many such systems of interest, there is 
a notable lack of powerful methods for computing the systems' Floquet states. 
Since these states constitute a natural basis which fully incorporates the 
periodic time dependence, thus providing a key for understanding both the 
short-time and the long-time behavior of periodically driven quantum systems, 
lack of knowledge of these states tends to limit one to a mere description of 
the results of numerical calculations, precluding deeper conceptual insight.

So far, numerical methods for treating large Floquet systems either make 
heavy use of specific properties of the respective system~\cite{MurilloEtAl15}
or require supercomputing facilities~\cite{LaptyevaEtAl16}, typically enabling
one to perform exact calculations for systems with a Hilbert space with a 
dimension on the order of $10^4$. In the present paper we establish a general 
computational strategy for calculating the Floquet states of even 
higher-dimensional systems with only modest numerical effort. This progress is 
achievable if one does not require all Floquet states and their quasienergies, 
but wishes to obtain information on specific individual states. In contrast 
to a variational principle which has been suggested recently for the same 
purpose~\cite{Krueger20}, here we propose an iterative algorithm that does 
not require a variational ansatz and can be executed whenever a sufficiently 
efficient scheme for propagating states in time is available. As an 
experiment-related application, we employ the exact Floquet states obtained 
in this manner for a periodically driven finite Bose-Hubbard chain in order to 
investigate the peculiarities of pseudo\-adiabatic Floquet-state preparation in 
the system's high-frequency regime~\cite{PolettiKollath11}. This is a somewhat 
subtle topic, since a proper adiabatic limit, referring to a turn-on of the 
driving amplitude that proceeds ``infinitely slowly'', cannot be expected to 
exist when the Bose-Hubbard system becomes large, owing to the fact that the 
gap condition required by the standard adiabatic theorem~\cite{BornFock28} 
cannot be satisfied; moreover, for an infinite system the quasienergy 
eigenvalues probably are nowhere differentiable with respect to the 
adiabatically changing parameter~\cite{HoneEtAl97,HoneEtAl09}. We argue that, 
nonetheless, there may be a window of opportunity involving driving amplitudes
which vary so fast that these pathologies remain almost unresolved, but still 
sufficiently slow to enable effectively adiabatic following at least to a 
high degree; the detailed experimental verification of this scenario with cold 
atoms in shaken optical lattices might constitute a rewarding challenge in the 
near future.   
   
We proceed as follows: In Sec.~\ref{S_2} we review some basic elements of 
the Floquet approach to periodically time-dependent quantum systems, thereby 
establishing our notation in the form it will be required later. We then 
introduce our iterative algorithm for computing Floquet states of 
high-dimensional systems in Sec.~\ref{S_3}, and discuss a method for 
``following'' Floquet states in response to small changes of the system's 
Hamiltonian in Sec.~\ref{S_4}. This type of computational following is still
not directly related to adiabatic following but based on the likeness of 
Floquet states. In Sec.~\ref{S_5} we apply these concepts to a periodically 
driven one-dimensional Bose-Hubbard chain, documenting both the viability of 
the iterative algorithm and the usefulness of the Floquet state-following 
scheme. In Sec.~\ref{S_6} we then tie open ends together and scrutinize what 
we call pseudoadiabatic following, demonstrating that a periodically driven 
many-body system may respond to a slowly changing driving amplitude by actually
taking the Floquet path sorted out by the computational following procedure.
Some conclusions regarding possible experimental implications are drawn in the 
final Sec.~\ref{S_7}.

\section{The Floquet picture}
\label{S_2}

Consider a quantum system with a periodically time-dependent Hamiltonian 
$\hat{H}(t) = \hat{H}(t+T)$ acting on a Hilbert space~${\mathcal H}$,
giving rise to the time-dependent Schr\"odinger equation 
\begin{equation}
	\ri\hbar\frac{\rd}{\rd t} | \psi(t) \rangle = 
	\hat{H}(t) | \psi(t) \rangle \; . 		
\label{eq:SEQ}
\end{equation}	
Then the unitary time-evolution operator of that system possesses the Floquet 
product form~\cite{Salzman74,BaroneEtAl77,GesztesyMitter81,Holthaus16}
\begin{equation}
	\hat{U}(t,0) = \hat{P}(t) \exp(-\ri \hat{G} t/\hbar) \; ,
\label{eq:REP}	
\end{equation}	
where the unitary operator $\hat{P}(t) = \hat{P}(t+T)$ inherits the periodic 
time dependence of the Hamiltonian, with $\hat{P}(0) = \hat{\mathds 1}$, and 
the  operator~$\hat{G}$ is self-adjoint. Thus, the transformed states
\begin{equation}
	| \widetilde{\psi}(t) \rangle = \hat{P}^\dagger(t) |\psi(t) \rangle
\end{equation}	
obey the Schr\"odinger equation
\begin{equation}
	\ri\hbar\frac{\rd}{\rd t} | \widetilde{\psi}(t) \rangle
	= \hat{G} | \widetilde{\psi}(t) \rangle \; ,
\label{eq:TRE}
\end{equation}
in which $\hat{G}$ appears as an effective time-independent Hamiltonian.

The quasienergies~$\varepsilon_n$ constitute the spectrum 
of~$\hat{G}$~\cite{GesztesyMitter81}. Requiring ${\mathcal H}$ to be of finite 
dimension, the ``stroboscopic'' approach to periodically time-dependent quantum
systems rests on the eigenvalue problem	
\begin{equation}
	\hat{U}(T,0) | n \rangle 
	= \exp(-\ri\varepsilon_n T/\hbar) | n \rangle
\end{equation}
which is posed by the unitary one-cycle evolution operator 
$\hat{U}(T,0) = \exp(-\ri \hat{G} T/\hbar)$ on ${\mathcal H}$. Since all 
eigenvalues $\exp(-\ri\varepsilon_n T/\hbar)$ fall on the unit circle of 
the complex plane, the quasienergies are thus determined up to an integer
multiple of $\hbar 2\pi/T \equiv \hbar\omega$.

If ${\mathcal H}$ is not of finite dimension, the eigenvalue problem may
become quite difficult. For example, the quasienergy spectrum of a linearly 
driven harmonic oscillator~\cite{PopovPerelomov70} is pure point if the driving
is nonresonant but absolutely continuous if the driving frequency matches 
the oscillation frequency of the undriven oscillator. More generally, the
question under which conditions the quasienergy spectrum of a periodically
driven quantum system may become continuous has been termed the ``quantum
stability problem'' in the literature~\cite{Combescure88}; some sophisticated 
theorems have been developed which allow one to exclude the presence of a
continuous spectrum in particular cases~\cite{Howland89,Howland92,Joye94}.
In order to avoid such complications we require ${\mathcal H}$ to be of large,
but finite dimension from here on.     

The ``extended'' viewpoint emerges when introducing the {\em Floquet 
functions\/}      
\begin{equation}
	 | u_n(t) \rangle = \hat{P}(t) | n \rangle
\end{equation}
which are $T$-periodic by construction,
\begin{equation}
	| u_n(t) \rangle = | u_n(t+T) \rangle \; .
\end{equation}
By virtue of the representation~(\ref{eq:REP}), the {\em Floquet states\/}
\begin{equation}
	| \psi_n(t) \rangle = | u_n(t) \rangle 
	\exp(-\ri\varepsilon_n t/\hbar)
\label{eq:FST}
\end{equation}		
then constitute a set of particular solutions to the Schr\"odinger 
equation~(\ref{eq:SEQ}) which is orthogonal and complete in ${\mathcal H}$
at each instant~$t$. Every solution to Eq.~(\ref{eq:SEQ}) can be expanded 
with respect to these states with time-independent coefficients~$a_n$,
\begin{equation}
	| \psi(t) \rangle = \sum_n a_n | u_n(t) \rangle 
	\exp(-\ri\varepsilon_n t/\hbar) \; ,
\end{equation}		
showing that the Floquet functions adopt a role which is conceptually similar
to that of the energy eigenfunctions of time-independent systems. Moreover,
inserting a Floquet state~(\ref{eq:FST}) into the Schr\"odinger equation, one 
readily derives
\begin{equation}
	\left( \hat{H}(t) + \frac{\hbar}{\ri} \frac{\rd}{\rd t} \right)
	| u_n(t) \rangle = \varepsilon_n | u_n(t) \rangle \; .
\label{eq:EVP}
\end{equation}
This is an eigenvalue equation for the quasienergies akin to the stationary
Schr\"odinger equation, posed by the quasienergy operator
\begin{equation}
	\hat{K} =  \hat{H}(t) + \frac{\hbar}{\ri} \frac{\rd}{\rd t}
\label{eq:KAM}
\end{equation}
on an extended Hilbert space $L_2[0,T] \otimes {\mathcal H}$ of
time-periodic functions~\cite{Sambe73}. Denoting the scalar product on 
${\mathcal H}$ by $\langle\, \cdot \, | \, \cdot \, \rangle$, the scalar
product on $L_2[0,T] \otimes {\mathcal H}$ is naturally given by
\begin{equation}
	\lF \, \cdot \, | \, \cdot \, \rF
	= \frac{1}{T} \int_0^T \! \rd t \, 
	\langle\, \cdot \, | \, \cdot \, \rangle \, ,
\label{eq:SCP}
\end{equation}  
since the time~$t$ is an additional coordinate in this extended space. When 
regarding a Floquet function $|u_n(t)\rangle$ no longer as a time-dependent
function on ${\mathcal H}$ but rather as an element of 
$L_2[0,T] \otimes {\mathcal H}$ it is written as $|u_n \rF$, so that the
quasienergy eigenvalue equation~(\ref{eq:EVP}) takes the form
\begin{equation}
	\hat{K} | u_n \rF = \varepsilon_n | u_n \rF \; .
\end{equation}
There is a seemingly simple but important implication that marks a crucial 
difference between this quasienergy eigenvalue problem and the more familiar 
energy eigenvalue problems encountered with time-independent Hamiltonian 
operators: If $|u_n(t)\rangle$ is a Floquet function with 
quasienergy~$\varepsilon_n$, and if~$m$ is any positive or negative integer, 
then $|u_n(t) \re^{\ri m\omega t}\rangle$ is a further $T$-periodic Floquet 
function with quasienergy $\varepsilon_n + m\hbar\omega$, where, again, 
$\omega = 2\pi/T$; all these different solutions are required for the 
completeness relation pertaining to the eigenfunctions of $\hat{K}$ in 
$L_2[0,T] \otimes {\mathcal H}$. On the other hand, solutions differing by~$m$ 
only give rise to the same Floquet state in ${\mathcal H}$, since
\begin{eqnarray} 
	& &
	|u_n(t) \re^{\ri m\omega t }\rangle 
	\exp\!\big(-\ri[\varepsilon_n + m \hbar\omega]t/\hbar\big) 
\nonumber \\	& = &	
	|u_n(t)\rangle\exp(-\ri \varepsilon_n t/\hbar) \; .	
\label{eq:INT}
\end{eqnarray}
Expressed pictorially, the spectrum of the quasienergy operator~(\ref{eq:KAM})
is obtained by ``unrolling'' the eigenvalues $\exp(-\ri\varepsilon_n T/\hbar)$
of the one-cycle evolution operator from the unit circle to the infinite real
energy axis. More technically, a ``quasienergy'' should not be regarded as a 
single eigenvalue, but rather as an infinite set of equivalent representatives 
spaced by $\hbar\omega$, implying that quasienergies cannot be ordered with 
respect to their magnitude without additional conventions. The spectrum of 
$\hat{K}$ thus consists of infinitely many identical Brillouin zones of width 
$\hbar\omega$, with each zone containing one quasienergy representative of 
each Floquet state. 

When the dimension of ${\mathcal H}$ becomes high, the distance between the 
quasienergy representatives within one Brillouin zone necessarily becomes 
small. On the other hand, being the eigenvalues of a Hermitian operator, 
the quasienergies are subject to the von Neumann-Wigner noncrossing rule 
and hence generally do not cross when only one parameter of the quasienergy 
operator is varied, unless such a crossing is permitted by a 
symmetry~\cite{NeumannWigner29}. Hence, excepting integrable systems, which 
actually do possess smooth quasienergies such as the nonresonantly driven 
harmonic oscillator~\cite{PopovPerelomov70}, the quasienergies of a 
high-dimensional periodically time-dependent quantum system form an intricate, 
dense net of avoided crossings when viewed as functions of one of its 
parameters, such as the amplitude of a driving force; the full complexity
of this net may remain unresolvable when limited to the accuracy attainable 
by numerical computations~\cite{HoneEtAl97,HoneEtAl09}. The question how to 
``follow'' an individual Floquet state on parameter changes in such a net 
is, therefore, far from trivial.

\section{Iterative computation of Floquet states}
\label{S_3}

Our present approach to computing individual Floquet states of periodically
time-dependent quantum systems is based on the following theorem, resorting
to the stroboscopic viewpoint:

Let $\hat{U}_1 = \hat{U}(T,0)$ be the one-cycle evolution operator of a 
quantum system defined on a finite-dimensional Hilbert space ${\mathcal H}$,
which possesses a time-periodic Hamiltonian $\hat{H}(t) = \hat{H}(t+T)$, and 
consider the functional $\Gamma_\gamma$ on ${\mathcal H}$ which is given by  
\begin{equation}
	\Gamma_\gamma [\,|\psi\rangle ] \equiv \langle \psi | 
	\left( \hat{U}_1 + \re^{-\ri\gamma} \right)^\dagger
	\left( \hat{U}_1 + \re^{-\ri\gamma} \right) | \psi \rangle \; .
\end{equation}
Then one has, for any $\gamma \in \mathbb R$ and any normalized 
$| \psi \rangle \in {\mathcal H}$,
\begin{equation}
	4 \ge \Gamma_\gamma [\,|\psi \rangle ] \ge 0 \; .
\end{equation}	
This is easily shown: After expanding $| \psi \rangle$ with respect to the 
eigenvectors $| n \rangle = |u_n(0) \rangle = |u_n(T) \rangle$ of $\hat{U}_1$,
obtaining
\begin{equation}
	| \psi \rangle = \sum_n a_n | n \rangle \; ,
\end{equation}
one finds
\begin{eqnarray}
	& & 	
	\left( \hat{U}_1 + \re^{-\ri\gamma} \right)^\dagger
	\left( \hat{U}_1 + \re^{-\ri\gamma} \right) | \psi \rangle
\nonumber \\	& = & \sum_n a_n 
	\Big( 2 + 2\cos(\varepsilon_n T/\hbar - \gamma) \Big) | n \rangle
\end{eqnarray}	
and, hence,
\begin{equation}
	\Gamma_\gamma [\,|\psi\rangle ] = \sum_n | a_n |^2
	\Big( 2 + 2\cos(\varepsilon_n T/\hbar - \gamma) \Big) \; . 
\end{equation}
From this representation the theorem follows immediately, since
$1 \ge \cos(\varepsilon_n T/\hbar - \gamma) \ge -1$, and $|\psi\rangle$ is
assumed to be normalized, $\sum_n | a_n |^2 = 1$. 
\hfill $\Box$

In particular, the maximum $\Gamma_\gamma [\,|\psi\rangle ] = 4$ of the 
functional $\Gamma_\gamma$ is attained if $|\psi \rangle = |n\rangle$ equals 
one of the eigenvectors of $\hat{U}_1$, and $\gamma = \varepsilon_n T/\hbar$
equals the corresponding phase. This observation enables one to invoke 
iterative methods for computing eigenvectors possessing the largest eigenvalue,
updating the phase~$\gamma$ at each step. Here we employ a power method based 
on the scheme
\begin{equation}  
	| \psi_{\rm new} \rangle = \left(
	\left( \hat{U}_1 + \re^{-\ri\gamma} \right)^\dagger
	\left( \hat{U}_1 + \re^{-\ri\gamma} \right) + \alpha
	\right) | \psi_{\rm old} \rangle \; ,	
\end{equation}
combined with subsequent normalization of $| \psi_{\rm new} \rangle$. The real 
parameter $\alpha$ can be adjusted empirically in order to speed up the 
convergence; evidently, one requires $\alpha > -2$ for filtering out 
the desired eigenvector of the kernel of the functional $\Gamma_\gamma$ 
which belongs to the largest eigenvalue. To be precise, we propose the 
following three-step algorithm:

{\em Step 1:\/} Choose a convenient initial state $| \psi_0 \rangle$ and 
propagate this state in time over one period~$T$, obtaining 
$\hat{U}_1|\psi_0\rangle$. If then
\begin{equation}
	1 - | \langle \psi_0 | \hat{U}_1 | \psi_0 \rangle | \le \delta \; ,
\label{eq:CON}
\end{equation} 	   
where $\delta > 0$ is a predefined small error tolerance, $|\psi_0\rangle$
already is an eigenvector $| n \rangle$ of $\hat{U}_1$, to the accuracy thus
specified. Hence, the algorithm terminates, and the corresponding quasienergy 
is obtained from the relation
\begin{equation}
	\langle \psi_0 | \hat{U}_1 | \psi_0 \rangle = 
	\exp(-\ri\varepsilon_n T/\hbar) \; .
\end{equation}
Otherwise, that is, if the condition~(\ref{eq:CON}) is not satisfied, compute
\begin{equation}
	\exp(-\ri\gamma_1) = 
	\frac{\langle \psi_0 | \hat{U}_1 | \psi_0 \rangle}
	{|\langle \psi_0 | \hat{U}_1 | \psi_0 \rangle|} \; ,
\end{equation}
yielding the vector
\begin{equation}
	| \psi_1 \rangle = 
	\left( \hat{U}_1 + \re^{-\ri\gamma_1} \right) | \psi_0 \rangle \; .
\end{equation}

{\em Step 2:\/} Now perform a propagation backward in time over one 
period~$T$ to obtain $\hat{U}_1^\dagger | \psi_1 \rangle$ and compute
\begin{equation}
	| \psi_2 \rangle = 
	\left( \hat{U}_1 + \re^{-\ri\gamma_1} \right)^\dagger 
	| \psi_1 \rangle \; .
\end{equation}

{\em Step 3:\/} Compute
\begin{equation}
	| \psi_3 \rangle = | \psi_2 \rangle + \alpha | \psi_0 \rangle
\label{eq:ALP}
\end{equation}
and normalize, obtaining
\begin{equation}
	| \psi_0 \rangle = | \psi_3 \rangle /
	\| \; | \psi_3 \rangle \; \| \; .
\end{equation}
With this new $|\psi_0\rangle$ go  back to Step~1 and repeat until the
algorithm terminates. 	  		
	 
Observe that this scheme for computing $|n\rangle = |u_n(0)\rangle$ and 
the quasienergy eigenvalue $\varepsilon_n$ can be executed already if a 
sufficiently efficient routine for propagating states in time is available. 
Thus, it is well applicable even if the dimension of ${\mathcal H}$ is so 
large that the computation and subsequent diagonalization of the one-cycle 
evolution operator, or alternatively the diagonalization of $\hat{K}$ in the 
extended Hilbert space, are rendered impracticable.

\section{Computational following of Floquet states}
\label{S_4}

Now suppose that a Floquet function $|u_1 \rF$ of some quasienergy operator
$\hat{K}_1$ with eigenvalue~$\varepsilon_1$ is already known, 
\begin{equation}
	\hat{K}_1 | u_1 \rF = \varepsilon_1 | u_1 \rF \; .
\label{eq:FSI}
\end{equation}
Next, suppose that the quasienergy operator is modified by adding a 
piece~$\lambda\hat{V}$, where $\lambda$ is a dimensionless para\-meter, giving
$\hat{K}_2 = \hat{K}_1 + \lambda\hat{V}$. It then appears natural to seed the 
algorithm devised in Sec.~\ref{S_3}, searching for a Floquet function of the 
new operator $\hat{K}_2$, with the old Floquet function $|u_1 \rF$ pertaining to
$\hat{K}_1$. This is based on the continuity assumption that for sufficiently 
small $\lambda$ there should be an eigenfunction of $\hat{K}_2$ which closely 
resembles $|u_1 \rF$. Would it then be possible to make a useful 
{\em a priori\/} statement concerning the performance of the algorithm?

To this end, consider the expression
\begin{equation}
	F[ \, |u\rF ; z; \hat{Y} ] \equiv \lF u | (\hat{Y} - z)^2 | u \rF \; ,
\label{eq:FUZ}
\end{equation}	
where $\hat{Y}$ is an operator acting on the extended Hilbert space, $|u \rF$ 
is an element of that space, $z$ is a scalar, and double angular brackets 
indicate the scalar product~(\ref{eq:SCP}) on $L_2[0,T] \otimes {\mathcal H}$.
In view of Eq.~(\ref{eq:FSI}), one evidently has
\begin{equation}
	F[ \, |u_1\rF ; \varepsilon_1; \hat{K}_1 ] = 0 \; ;
\end{equation}
this identity implies a variational principle for Floquet 
states~\cite{Krueger20}. Inserting the old, known Floquet function~$|u_1\rF$ 
and its quasienergy~$\varepsilon_1$, but the new  quasienergy operator
$\hat{K}_2$ into this expression~(\ref{eq:FUZ}), it jumps to the nonzero value
\begin{equation}
	F[ \, |u_1\rF ; \varepsilon_1; \hat{K}_2 ] = 
	\lF u_1 | (\lambda\hat{V})^2 | u_1 \rF \; .
\end{equation}
More elaborately, computing
\begin{eqnarray}
	\varepsilon & = & \lF u_1 | \hat{K}_2 | u_1 \rF
\nonumber \\	& = &	
	\varepsilon_1 + \lF u_1 | \lambda\hat{V} | u_1 \rF \; , 	
\end{eqnarray}
and inserting this $\varepsilon$ instead of $\varepsilon_1$, one derives
\begin{eqnarray}
	F[ \, |u_1\rF ; \varepsilon; \hat{K}_2 ] & = & \lF u_1 | 
	\left( \lambda\hat{V} - \lF u_1 | \lambda\hat{V} | u_1 \rF \right)^2 
	| u_1 \rF
\nonumber \\	& \equiv &
	{\rm Var}_1(\lambda\hat{V}) \; .
\label{eq:VAR}
\end{eqnarray}	
Under the plausible, yet unproven assumption that the magnitude of this jump 
of $F$ from zero to ${\rm Var}_1(\lambda\hat{V})$ is monotonically related to 
the number of iterations it takes the algorithm to converge to the new Floquet 
state, one obtains a cue how to choose the parameter $\lambda$. In particular, 
when trying to monitor the behavior of a certain Floquet state for all values 
of~$\lambda$ within some interval $[0, \lambda_{\rm max}]$ of interest by 
performing computations on a grid of width $\Delta\lambda$, it may be helpful 
to compute ${\rm Var}_1(\hat{V})$ at each step, thus obtaining information on 
``how far away'' the desired new Floquet state may be from the previous one; 
if this jump appears too large, $\Delta\lambda$ should be suitably decreased.

We emphasize that this strategy for following an individual Floquet state in 
parameter space does not necessarily result in following by continuity with 
respect to~$\lambda$. If, for instance, the relevant quasienergy functions 
$\varepsilon_n(\lambda)$ are broken by a large number of tiny avoided 
crossings, indicating weak multiphoton-like resonances, and if the increment 
$\Delta\lambda$ is larger than the typical width of these avoided crossings, 
the initial state will be followed diabatically, that is, ignoring the avoided 
crossings as if they did not exist. If, on the other hand, $\Delta\lambda$ is 
comparable to the size of the avoided crossings, they will be resolved, and the 
numerically computed Floquet state will follow its quasienergy continuously, 
that is, adiabatically. It remains to be explored whether this potential 
sensitivity of the computational following procedure to the stepsize 
$\Delta\lambda$ can be exploited for extracting useful information about the 
physics of the respective system under investigation.

\section{Application: The periodically driven Bose-Hubbard model}
\label{S_5}

The Bose-Hubbard model constitutes an idealized lattice system of theoretical 
many-body physics, embodying the competition between delocalization due 
to kinetic energy and localization due to repulsive potential energy, 
thus giving rise to a superfluid-Mott insulator quantum phase 
transition~\cite{FisherEtAl89}. Here we consider a one-dimensional 
Bose-Hubbard chain, as specified by the Hamiltonian
\begin{equation}
	\hat{H}_0 = \hat{H}_{\rm tun} + \hat{H}_{\rm int}
\label{eq:HBH}
\end{equation}
with nearest-neighbor tunneling
\begin{equation}
	\hat{H}_{\rm tun} = -J \sum_j \left( 
	\hat{a}_j^\dagger     \hat{a}_{j+1}^{\phantom\dagger} +
	\hat{a}_{j+1}^\dagger \hat{a}_j^{\phantom\dagger} \right)
\end{equation}
and on-site interaction
\begin{equation}
	\hat{H}_{\rm int} = 
	\frac{U}{2} \sum_j \hat{n}_j \left( \hat{n}_j - 1 \right) \; ,  
\end{equation}	
where $j$ labels the chain's sites in consecutive order, 
$\hat{a}_j^{\phantom\dagger}$ ($\hat{a}_j^\dagger$) annihilates (creates) 
a Bose particle at site~$j$, implying the canonical commutator 
$[ \hat{a}_j^{\phantom\dagger} , \hat{a}_k^\dagger] = \delta_{jk}$, while
$\hat{n}_j = \hat{a}_j^\dagger \hat{a}_j^{\phantom\dagger}$ denotes the 
number operator at that site. Moreover, $J$ is the nearest-neighbor hopping 
matrix element, and $U$ is the repulsion energy contributed by each pair 
of particles occupying a common site. In the limit of infinite chain length, 
the phase transition occurs at $(J/U)_{\rm c} = 0.297 \pm 0.01$ for unit 
filling, that is, when the chain is occupied by one particle per 
site~\cite{KuehnerMonien98,KuehnerEtAl00}.  

In order to equip this model with a periodic time dependence, thus admitting
an additional wealth of Floquet physics, we introduce a monochromatic 
homo\-geneous driving force described by 
\begin{equation}
	\hat{H}_{\rm drive}(t) = R \cos(\omega t) \sum_j j \hat{n}_j \; ,
\label{eq:HDR}
\end{equation}	
and investigate the periodically driven Bose-Hubbard model~\cite{EckardtEtAl05} 
\begin{equation}
	\hat{H}(t) = \hat{H}_0 + \hat{H}_{\rm drive}(t) \; ;
\label{eq:DBH}
\end{equation}	
this driving scheme can be implemented experimentally with ultracold atoms in 
periodically shaken optical lattices~\cite{DreseHolthaus97,EckardtEtAl05, 
ZenesiniEtAl09,ArimondoEtAl12}. Some salient features of this 
system~(\ref{eq:DBH}) can already be deduced by calculating the matrix 
elements of its quasienergy operator~$\hat{K}$ for the Floquet-Fock 
functions~\cite{EckardtEtAl05} 
\begin{eqnarray} 
	& & | \{n_j\} , m \rF 
\nonumber\\	& = &	
	| \{ n_j \} \rangle
	\exp\!\left( - \ri \frac{R}{\hbar\omega} \sin(\omega t)
	\sum_j j n_j + \ri m \omega t \right)
\end{eqnarray}
spanning its extended Hilbert space, where $| \{ n_j \} \rangle$ denotes a 
Fock state with $n_j$ particles occupying the site labeled by~$j$, and $m$ 
is the integer having shown up before in Eq.~(\ref{eq:INT}): While the on-site 
contributions remain diagonal,
\begin{eqnarray}
	& &
	\lF \{n'_j\} , m' | \hat{H}_{\rm int} + H_{\rm drive}(t) 
	+ \frac{\hbar}{\ri} \frac{\rd}{\rd t} | \{n_j\} , m  \rF
\nonumber \\ & = &
	\left[ \frac{U}{2} \sum_j n_j (n_j - 1) + m\hbar\omega \right]
	\delta_{\{n_j\},\{n'_j\}} \delta_{m,m'} \; ,	
\end{eqnarray}	 
the matrix elements of the tunneling operator are found to read
\begin{eqnarray}
	& &
	\lF \{n'_j\} , m' | \hat{H}_{\rm tun} | \{n_j\} , m  \rF = 	
	\langle \{n'_j\} | \hat{H}_{\rm tun} | \{n_j\} \rangle 
\nonumber \\	& \times &	
	\frac{1}{T}\!\int_0^T \!\! \rd t \, \exp\!\left(
	-\ri\frac{Rs}{\hbar\omega}\sin(\omega t) + \ri(m - m')\omega t 
	\right) \, ,	
\end{eqnarray}
where 
\begin{equation}
	s = \sum_j j (n_j - n'_j) = \pm 1 \; ,	
\end{equation}
since $\hat{H}_{\rm tun}$ transfers one particle by one site along the chain, 
with the plus (minus) sign referring to a tunneling process to the left 
(right). Invoking the Jacobi-Anger identity
\begin{equation}
	\exp(\ri z \sin\varphi) = \sum_{\ell=-\infty}^{+\infty}
	\re^{\ri\ell\varphi} {\mathcal J}_\ell(z) 
\end{equation}
for the Bessel functions ${\mathcal J}_\ell$ of integer order~$\ell$, 
this becomes 
\begin{eqnarray}
	& &
	\lF \{n'_j\} , m' | \hat{H}_{\rm tun} | \{n_j\} , m  \rF 
\nonumber \\ 	& = &		
	\langle \{n'_j\} | \hat{H}_{\rm tun} | \{n_j\} \rangle \, (-s)^{m'-m} 
	\, {\mathcal J}_{m'-m}\!\left( \frac{R}{\hbar\omega} \right) \; .
\end{eqnarray}	
In the high-frequency regime, in which the width $\hbar\omega$ of the 
quasienergy Brillouin zone becomes the dominant energy scale and ``small''
avoided crossings of quasienergy representatives belonging to different~$m$
can be ignored, one may neglect all Bessel function factors 
${\mathcal J}_{m'-m}(R/\hbar\omega)$, except for $m'-m = 0$. Thus, one arrives 
at a  time-independent effective Hamiltonian which differs from the original, 
undriven Bose-Hubbard model~(\ref{eq:HBH}) only through the replacement 
of the hopping matrix element~$J$ by the ``renormalized'' hopping 
strength~\cite{EckardtEtAl05}
\begin{equation}
	J_{\rm eff} = J \, {\mathcal J}_0(R/\hbar\omega) \; ;
\label{eq:JEF}
\end{equation}
this effective Hamiltonian is a simple, but often apparently sufficient 
approximation to the exact operator~$\hat{G}$ introduced on general grounds in 
Eqs.~(\ref{eq:REP}) and~(\ref{eq:TRE}). Since the iterative scheme established 
in Sec.~\ref{S_3} does not require any such approximation, we are now in a 
position to subject this high-frequency approximation to a ``hard'' numerical 
test.

The interest in such a test stems from the following deliberation: 
Assume that $J/U > (J/U)_{\rm c}$, so that the undriven chain~(\ref{eq:HBH}) 
is in its superfluid ground state. If then the scaled driving amplitude 
$R/(\hbar\omega)$ is increased from zero toward the first zero 
$j_{0,1} \approx 2.405$ of the Bessel function ${\mathcal J}_0$, the effective 
hopping strength~(\ref{eq:JEF}) decreases monotonically to zero. This implies 
that the effective time-independent model enters the Mott insulator regime at a 
certain ``critical'' driving strength, at which $J_{\rm eff}/U = (J/U)_{\rm c}$;
for even stronger driving, the energy ground state of the effective model is 
separated from the excited states by a finite gap. Thus, the effective 
model predicts the emergence of a driving-induced Mott insulator 
state~\cite{EckardtEtAl05}.

The full Floquet system~(\ref{eq:DBH}), however, does not possess such a 
gapped ground state. Considering an infinitely long chain for the sake of 
the argument, the quasienergy of a Floquet state originating on activation 
of the drive from the undriven chain's Mott insulator ground state would be 
embedded in the continuum of excited states due to the Brillouin zone 
structure of the quasienergy spectrum. Therefore, this state should turn into 
a resonance, that is, into a decaying state characterized by a Lorentzian 
peak in the spectral density with a certain width determining its life time, 
akin to the familiar Floquet resonances of atomic states in strong laser
fields~\cite{Yajima82}, so that the decay width may be interpreted as an
inverse characteristic heating time.

Returning to large but finite driven Bose-Hubbard chains which are more 
realistic from an experimental point of view, the quasienergy continuum 
resulting from the excited states gives way to a huge number of discrete
quasienergy eigenvalues filling each Brillouin zone. Thus, on variation of 
the driving amplitude, a Floquet state associated with a Mott insulator 
state would not be protected by a gap but would have to avoid a plethora 
of other states instead, to the effect that its quasienergy cannot depend 
smoothly on the driving amplitude, but is ``broken'' by countless tiny 
avoided crossings. Indeed this ``roughness'' of the quasienergy functions 
should enable one to estimate the corresponding decay or heating times, 
by a procedure mimicking the so-called ${\mathcal L}^2$ stabilization
method~\cite{MandelshtamEtAl93,MakarovEtAl94}. Here we do not attempt to
determine these heating times, but focus on the first and foremost question:  
If one goes beyond the convenient, but essentially uncontrolled high-frequency
approximation~(\ref{eq:JEF}), would it actually be possible to identify 
driving-induced Mott insulatorlike Floquet states despite the above caveats? 

To this end we study a finite Bose-Hubbard chain~(\ref{eq:HBH}) possessing
$M = 11$ sites ranging from $j_{\rm min} = -5$ to $j_{\rm max} = 5$, and 
specify unit filling, so that the chain is occupied by $N = 11$ Bose particles.
The dimension~$d$ of this system's Hilbert space ${\mathcal H}$ then is 
determined by the binomial coefficient
\begin{equation}
	d = \left( \! \begin{array}{c} N+M-1 \\ M-1 \end{array} \! \right)
	= 352716 \; ,
\end{equation}
posing a serious challenge to more traditional methods commonly used for 
investigating Floquet systems. Adopting the above reasoning, we fix the 
parameter $J/U = 1/3 > (J/U)_{\rm c}$, placing the undriven system in the 
superfluid regime. The energy spectrum of this chain ranges from 
$E_{\min}/U = -3.64$ to $E_{\max}/U = 55.25$. Moreover, we select the driving 
frequency $\hbar\omega/U = 14/3$, thus aiming for the high-frequency regime in 
which the approximation~(\ref{eq:JEF}) should be viable~\cite{EckardtEtAl05}. 
This gives $E_{\min}/(\hbar\omega)= -0.78$ and $E_{\max}/(\hbar\omega) = 11.84$,
implying  that the chain's energy spectrum covers more than~$12$ quasienergy 
Brillouin zones. With this choice of para\-meters one finds
$J_{\rm eff}/U = (J/U)_{\rm c}$ for $R/(\hbar\omega) \approx 0.67$. Hence
the driven system should enter a Mott insulatorlike regime at about this 
value of the the driving amplitude, and stay therein for stronger driving. 

The quasienergy operator $\hat{K}$ of the driven chain remains invariant 
under a generalized parity transformation in its extended Hilbert space, 
corresponding to the spatial reflection $j \to -j$ combined with a shift in 
time by half a period, $t \to t + T/2$. Since the Floquet functions are odd 
or even under this generalized parity, and eigenvalues belonging to the same 
parity class are subject to the von Neumann-Wigner noncrossing rule, each of 
the $d$~quasienergy respresentatives falling into one Brillouin zone inevitably 
is perforated by a large number of avoided crossings when, for example, the 
driving amplitude is varied.

\begin{figure}[t]
\centering
\includegraphics[width=0.9\linewidth]{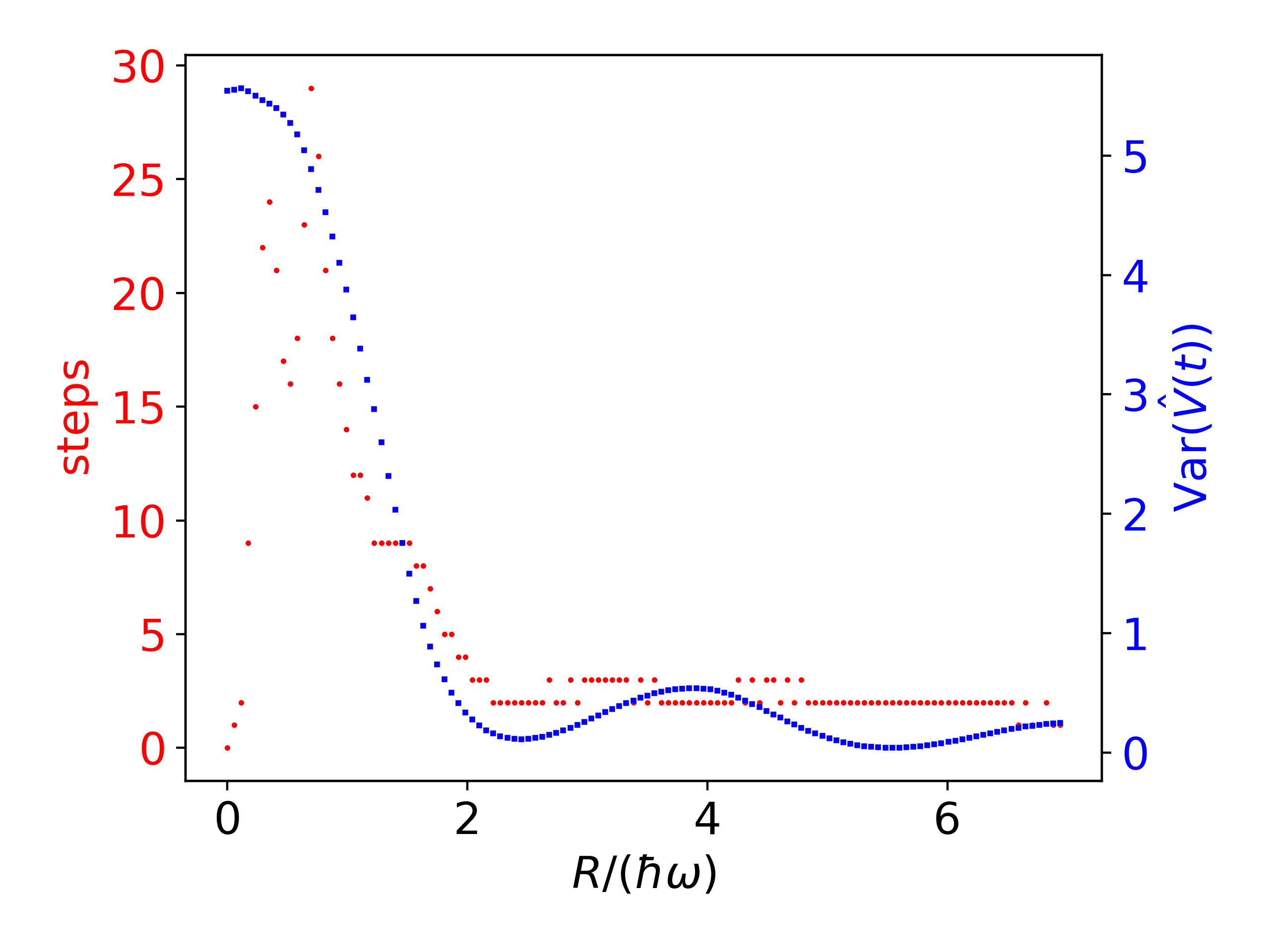}
\caption{Following the Floquet state that develops from the energy ground 
	state of the Bose-Hubbard chain~(\ref{eq:HBH}) with $M = 11$ sites
	occupied by $N = 11$ particles in response to the drive~(\ref{eq:HDR})
	with parameters $J/U = 1/3$ and $\hbar\omega/U = 14/3$: Red dots 
	indicate the number of iterations required for converging to the 
	respective new Floquet state for stepsize 
	$\Delta R/(\hbar\omega) = 7/120$, tolerance $\delta = 10^{-4}$, and 
	$\alpha = -1.9$; blue dots show the magnitude of the corresponding 
	jump ${\rm Var}_1(\hat{V}(t))$, where 
	$\hat{V}(t) = \hat{H}_{\rm drive}(t)/R$.} 	   
\label{F_1}
\end{figure}

As a preliminary application of the strategy put forward in Sec.~\ref{S_4}, 
we try to follow the Floquet state which develops from the superfluid 
energy ground state of the undriven chain in response to a time-periodic 
drive~(\ref{eq:HDR}) with increasing driving amplitude, so that the scaled 
amplitude~$R/(\hbar\omega)$ now plays the role of the parameter~$\lambda$. 
The calculations were performed on a standard laptop computer equipped
with a GPU (GeForce GTX 1060 Mobile).
Figure~\ref{F_1} depicts the performance of the algorithm when scanning the 
interval $0 \le R/(\hbar\omega) \le 7$ with stepsize $\Delta R/(\hbar\omega) = 
7/120$ and tolerance $\delta = 10^{-4}$; here, the parameter~$\alpha$ employed 
in Eq.~(\ref{eq:ALP}) has been set to $\alpha = -1.9$. These results are fairly 
encouraging: Indeed, the procedure always converges to a Floquet state after 
less than 30 iterations. Most significantly, the expected entrance into a 
Mott insulatorlike regime is reflected by the number of iterations; in the 
initial superfluidlike regime, the state is increasingly harder to follow 
when $R/(\hbar\omega)$ is enhanced. This behavior changes after entering the 
potential Mott regime; for $R/(\hbar\omega) > 2$, very few iterations 
are required for settling down to the new Floquet state. Although there can 
be no sharp transition here, related features are observed when monitoring 
the functional ${\rm Var}_1(\hat{H}_{\rm drive}(t)/R)$, as defined by 
Eq.~(\ref{eq:VAR}). In particular, the low values of this functional for 
large driving amplitudes appear to match the incompressibility which is 
characteristic of a genuine Mott insulator~\cite{FisherEtAl89}, since the 
system's compressibility is determined by the variances of the particle-number 
operators showing up in $\hat{H}_{\rm drive}(t)$.

\begin{figure}[t]
\centering
\includegraphics[width=0.9\linewidth]{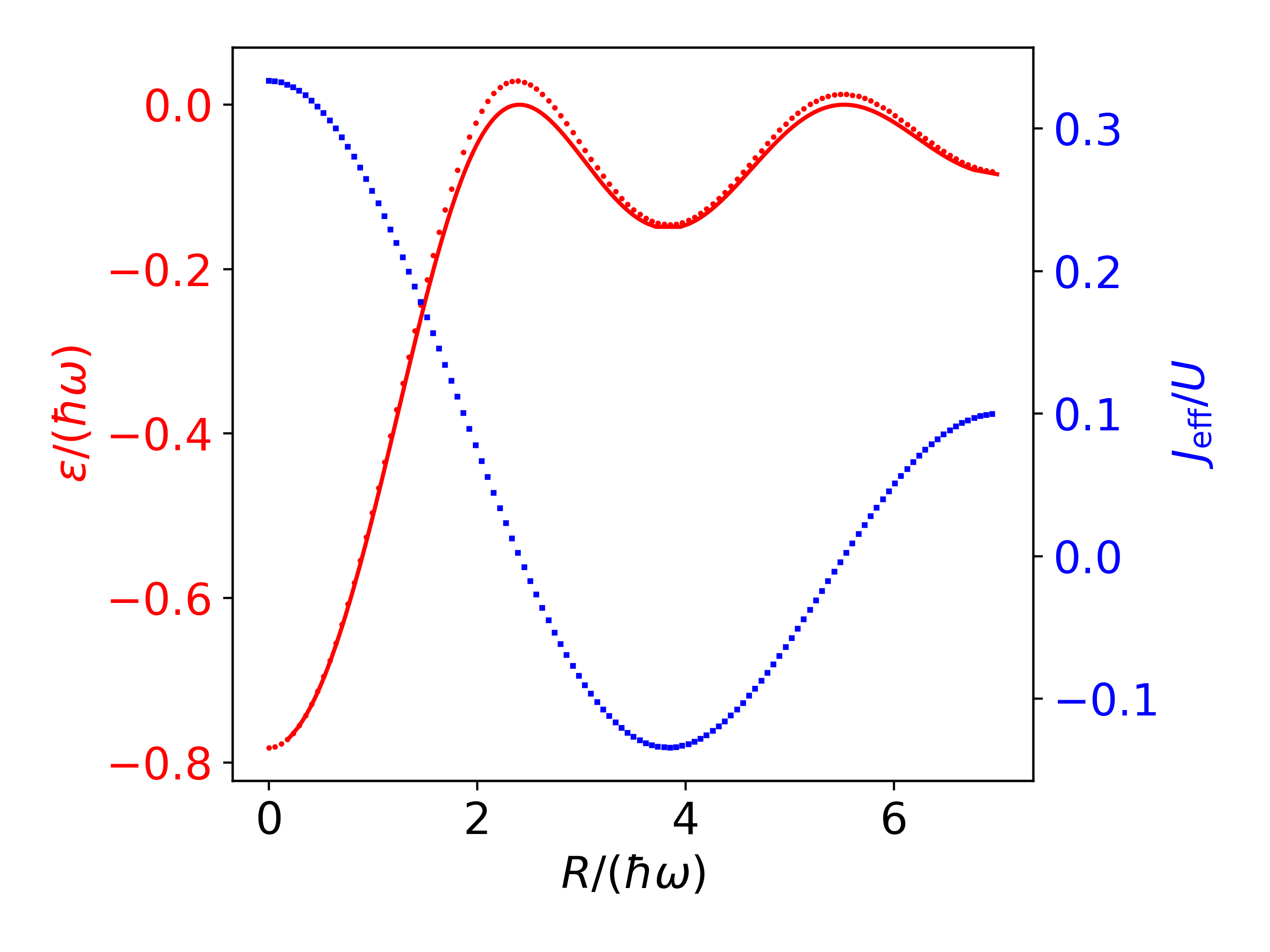}
\caption{One representative of the quasienergy of the Floquet state resulting
	from the numerical following scheme documented in Fig.~\ref{F_1}
	(red dots) compared with the corresponding ground-state energy
	of the effective Hamiltonian within the high-frequency approximation
	(red full line). Blue dots indicate the effective hopping 
	strength~(\ref{eq:JEF}). Observe that the quasienergy becomes maximal
	when this effective hopping strength vanishes.}   
\label{F_2}
\end{figure}

In Fig.~\ref{F_2} we compare the quasienergies of the exact, numerically 
computed Floquet states obtained by this following procedure to the 
corresponding ground state energy of the effective Hamiltonian with the 
appropriately renormalized hopping strength~(\ref{eq:JEF}). The agreement is 
quite impressive except for scaled driving amplitudes $R/(\hbar\omega)$ close 
to the zeros $j_{0,1} \approx 2.405$ and $j_{0,2} \approx 5.520$ of the Bessel 
function ${\mathcal J}_0$: At these zeros, the ground-state energy of the 
effective time-independent model vanishes, while the exact quasienergies rise 
to slightly higher values. Still, it needs to be kept in mind that the exact 
Floquet state considered here is coupled to a background of many others by 
large numbers of avoided crossings; yet, these avoided crossings are too 
narrow to be individually resolved on the scale imposed in Fig.~\ref{F_2}. 
In contrast, the energy of the effective model's ground state varies in a 
perfectly smooth manner with the driving amplitude. Under the present 
conditions the Floquet state following procedure thus provides a coarse-grained 
image of the exact eigenvalue, selecting the followers diabatically, that is, 
according to maximum likeness of the states. The fact that no ``roughness'' 
of the quasienergy can be detected in Fig.~\ref{F_2} also implies that the 
corresponding Floquet resonance is fairly long-lived.

\begin{figure}[t]
\centering
\includegraphics[width=0.9\linewidth]{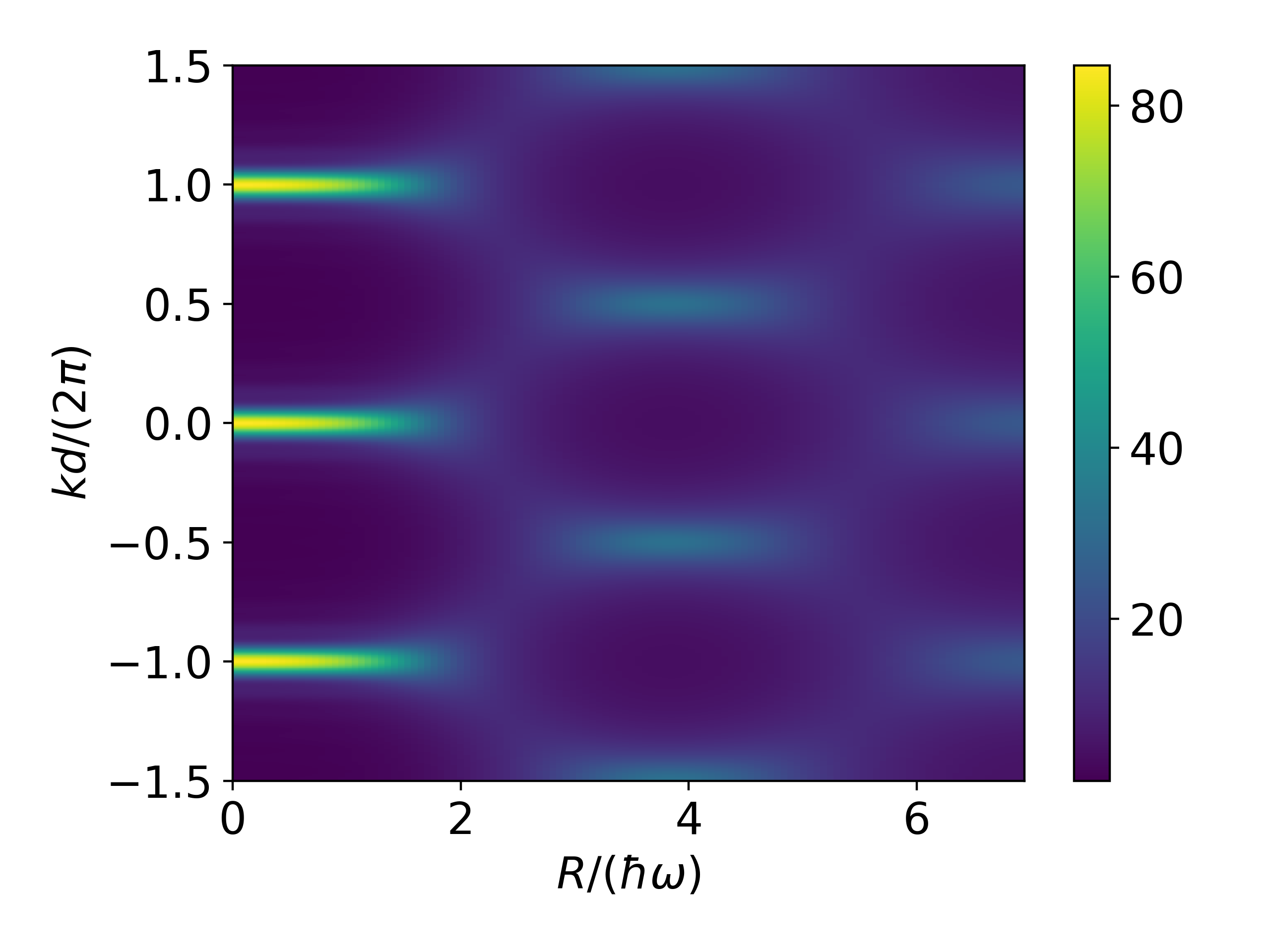}
\caption{Contour plot of the momentum distribution functions~(\ref{eq:MDF}) 
	pertaining to the effective model. For scaled driving amplitudes 
	$R/(\hbar\omega)$ close to the zeros $j_{0,1} \approx 2.405$ and 
	$j_{0,2} \approx 5.520$ of ${\mathcal J}_0$, this function is almost 
	structureless; between these zeros, one observes a weak precursor of 
	a new superfluid state.}  
\label{F_3}
\end{figure}

On the level of the states, the fair agreement between the effective and 
the full, periodically time-dependent system in the high-frequency regime 
is emphasized further by the following comparison. Figure~\ref{F_3} shows 
contour plots of the momentum distribution functions   
\begin{equation}
	C_{\rm eff}(k) = \sum_{j,l} \langle \varphi_0 |
	\hat{a}_j^\dagger \hat{a}_l^{\phantom\dagger} 
	| \varphi_0 \rangle \exp\!\big(-\ri k [j-l]d\big)
\label{eq:MDF}	
\end{equation}
of the effective model, where~$k$ is a wave number and $d$ indicates the 
lattice constant; $| \varphi_0 \rangle$ is the energy ground state of the 
effective model at the respective value of $R/(\hbar\omega)$. Such momentum
distributions are accessible to measurement with the help of time-of-flight 
interference experiments~\cite{GreinerEtAl02,BlochEtAl08,HoffmannPelster09}. 
One observes marked peaks of this distribution function at $k = 0 \bmod 2\pi/d$
in the superfluid regime at low driving amplitudes, since the system tends to 
condense into the single-particle state at the bottom of the energy band 
\begin{equation}
	E(k) = -2J \cos(kd)
\label{eq:BND}	
\end{equation}	
of the noninteracting system with $U = 0$. For scaled driving amplitudes
$R/(\hbar\omega)$ in the vicinity of the zeros $j_{0,1}$ and $j_{0,2}$ of 
${\mathcal J}_0$ this distribution function is almost flat, as it should, 
since the effective hopping matrix element~(\ref{eq:JEF}) is small there. 
Interestingly, between these zeros one finds a weak precursor of a new 
superfluid state conforming to a condensate in the single-particle state 
$k = \pi/d \bmod 2\pi/d$, as corresponding to the upper edge of the 
band~(\ref{eq:BND}); this is naturally explained by the fact that the sign 
of the effective hopping strength is inverted between the zeros.

\begin{figure}[t]
\centering
\includegraphics[width=0.9\linewidth]{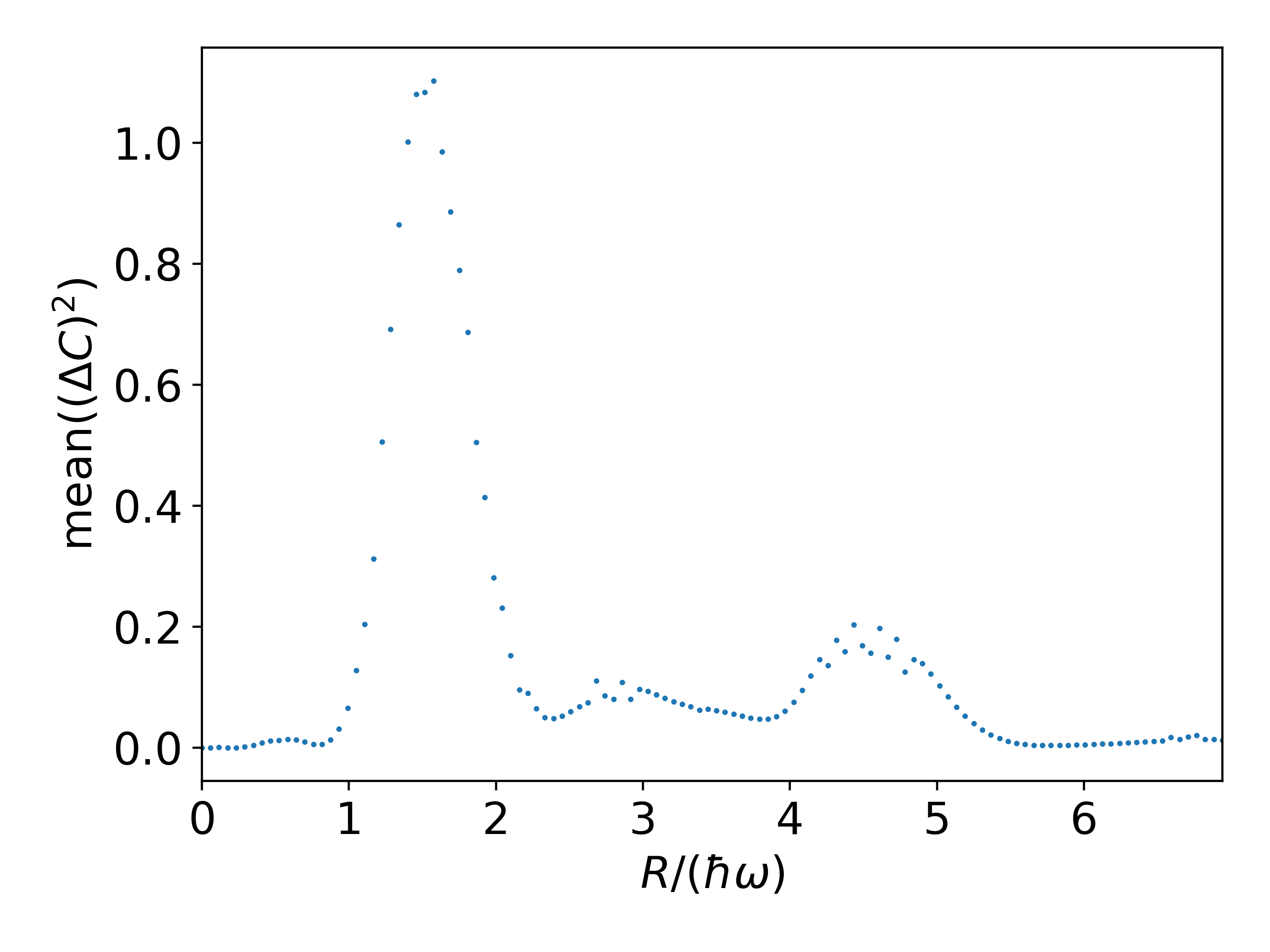}
\caption{The squared difference $C_{\rm eff}(k) - C_{\rm exact}(k)$ of the
	momentum distribution function~(\ref{eq:MDF}) predicted by the
	effective model and that obtained for the exact Floquet states
	computed by the following scheme, averaged over~$k$. Note how the
	scale here differs from that in Fig.~\ref{F_3}.}
\label{F_4}
\end{figure}

When the energy eigenstates $| \varphi_0 \rangle$ entering the 
functions~(\ref{eq:MDF}) are replaced by the exact Floquet functions 
$| u(0) \rangle$ computed by following the ground state of the undriven
chain with increasing driving amplitude, providing the exact distribution
functions $C_{\rm exact}(k)$, one obtains a plot that is almost 
indistinguishable from Fig.~\ref{F_3}. Therefore, we depict in Fig.~\ref{F_4} 
the squared difference 
\begin{equation}
	{\rm mean}_k \left((\Delta C)^2\right) =
	{\rm mean}_k \left( [C_{\rm eff}(k) - C_{\rm exact}(k)]^2 \right) \, 
\end{equation}
averaged over~$k$, revealing that the deviation between the two functions 
figures on the subpercent level. Thus, under the conditions considered in 
this section, the simple time-independent, effective model performs fairly 
well when compared with the full, periodically driven Bose-Hubbard 
chain~(\ref{eq:DBH}), and the question posed above can be answered in the 
affirmative: The driving-induced Mott insulatorlike Floquet state predicted 
by the high-frequency approximation~(\ref{eq:JEF}) is no artifact.

\section{Pseudoadiabatic following of Floquet states}
\label{S_6}

Going beyond the computational Floquet state-following strategy suggested 
in Sec.~\ref{S_4} and practiced for one particular example in Sec.~\ref{S_5}, 
there also is the more experiment-related question of whether the 
actual state of a system would be able to follow one of its Floquet states in 
an adiabatic manner when its parameters are changing in real time. Indeed
there is an adiabatic principle for Floquet states~\cite{BreuerHolthaus89a,
DreseHolthaus99}, which formally resembles the celebrated adiabatic theorem of 
quantum mechanics~\cite{BornFock28}. This principle allows one, for instance, 
to define generalized $\pi$-pulses together with a generalized area theorem, 
which govern transitions in periodically driven multilevel ladder systems in 
close analogy to their two-level analogs~\cite{HolthausJust94}. However, for 
many-body Floquet systems with a high-dimensional Hilbert space, as considered 
here, the inevitable multitude of avoided crossings renders adiabatic 
following, if understood in the formal mathematical sense, impossible. Yet, a 
closely related option emerges: If the relevant quasienergies are punctured 
only by narrow avoided crossings below a certain scale, and if the parameter 
variation, while sufficiently slow, still proceeds so fast that these narrow 
avoided crossings are traversed almost diabatically by Landau-Zener-like 
Floquet-state transitions~\cite{BreuerHolthaus89b}, highly diabatic quantum 
motion actually appears as effectively adiabatic motion on coarse-grained 
quasienergy eigenvalue surfaces~\cite{BreuerHolthaus89a,Holthaus16}. This is termed 
pseudoadiabatic following here, enabling one to design the parameter 
variation such that desired Floquet target states are populated with high 
probability. 
      
For demonstrating the feasibility of this concept we now perform a series 
of numerical experiments: Initially, at time $t = 0$, the state 
$| \psi(t=0) \rangle$ of the system is given by the energy ground state of 
the undriven Bose-Hubbard chain~(\ref{eq:HBH}). Then the driving amplitude 
is smoothly ramped up according to the protocol
\begin{equation}
	R(t) = R_{\max} \sin^2\!\left( \frac{t}{zT} \frac{\pi}{2} \right)
	\quad \text{for } \; 0 \le t \le zT \; , 
\label{eq:RMP}
\end{equation}
reaching its final value~$R_{\rm max}$ at time $t = zT$, that is, after~$z$
driving cycles $T = 2\pi/\omega$. The resulting state $|\psi(t = zT) \rangle$, 
as computed by numerical integration of the Schr\"odinger equation, is then 
projected on the Floquet function $| u(0) \rangle$ calculated for $R = R_{\max}$
by means of the procedure outlined in Sec.~\ref{S_5}, yielding the overlap
\begin{equation}
	O = \langle u(0) | \psi(zT) \rangle \; .
\label{eq:OVL}
\end{equation}   
Similar studies have been reported by Poletti and Kollath in 
Ref.~\cite{PolettiKollath11}, the difference being that here we target an 
exact Floquet state instead of the ground state of the effective Hamiltonian.

\begin{figure}[t]
\centering
\includegraphics[width=0.9\linewidth]{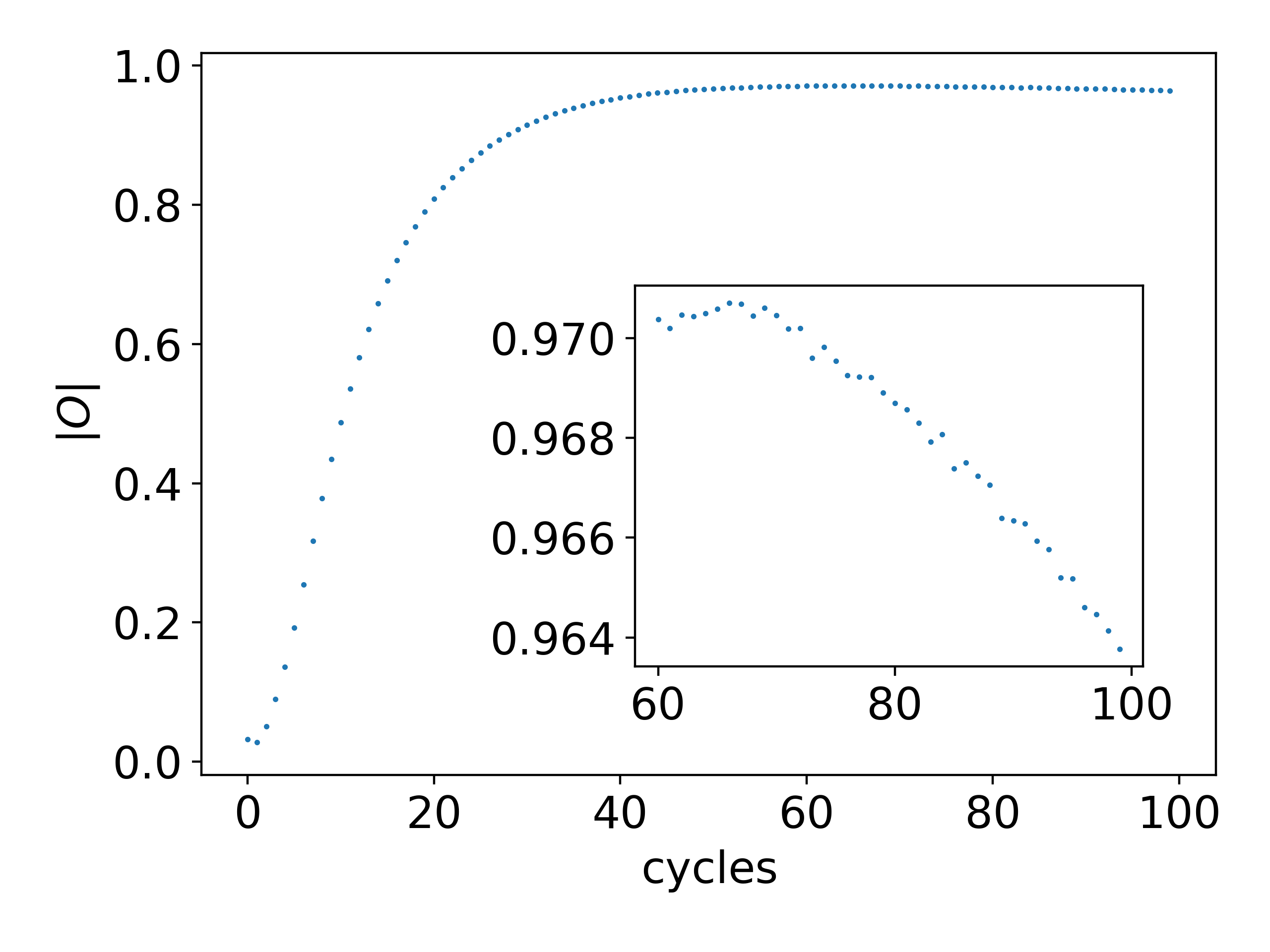}
\caption{Absolute value of the overlap~(\ref{eq:OVL}) resulting from a smooth 
	ramp~(\ref{eq:RMP}) of the driving amplitude from zero to the final 
	value~ $R_{\max}/(\hbar\omega) = 4$, vs the number~$z$ of driving
	cycles it takes to reach this final amplitude. Parameters are
	$J/U = 1/3$ and $\hbar\omega/U = 14/3$, as corresponding to
	Figs.~\ref{F_1}	and~\ref{F_2}.}    
\label{F_5}
\end{figure}

Figure~\ref{F_5} depicts numerical results obtained in this manner for 
$R_{\max}/(\hbar\omega) = 4$, again with $J/U = 1/3$ and $\hbar\omega/U = 14/3$:
While the absolute value of the overlap~(\ref{eq:OVL}) is only small if the 
turn-on of the driving force proceeds within a few cycles, because the initial 
superfluid state is given insufficient time to adjust to the Mott insulatorlike
target state, it becomes appreciably higher than~$0.95$ when the final amplitude
is reached after some $10$ cycles. Yet, the inset of Fig.~\ref{F_5} reveals 
that this trend cannot be extended to much longer turn-on times: For $z > 70$
the population of the target Floquet state starts to decrease. This is the
expected consequence of the many tiny avoided crossings that have remained 
unresolved in Fig.~\ref{F_2} but nonetheless are effective: Even if each 
individual avoided crossing is still traversed almost diabatically when~$z$ 
is increased, they conspire by their sheer number to divert an increasing 
fraction of the evolving state into the anticrossing Floquet states; this 
fraction cannot reach the target state. This is, in a nutshell, a 
visualization of the window of opportunity referred to in the Introduction: 
Pseudoadiabatic following of Floquet states is possible if {\em (i)\/} their 
quasienergies do not exhibit large avoided crossings, and {\em (ii)\/} the 
parameter variation proceeds reasonably slowly, but not too slowly.

\begin{figure}[t]
\centering
\includegraphics[width=0.9\linewidth]{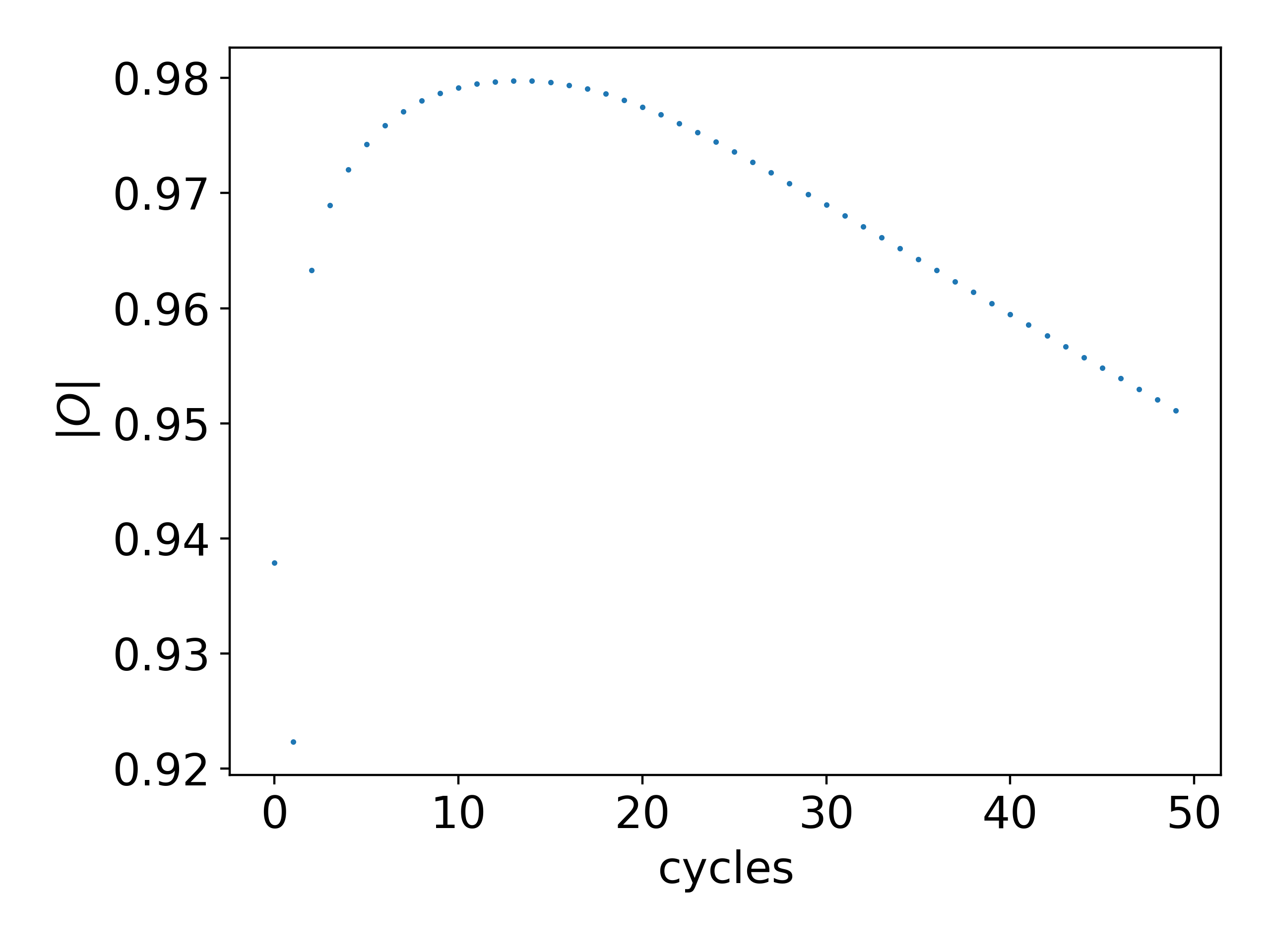}
\caption{As Fig.~\ref{F_5} but for lower value $R_{\max}/(\hbar\omega) = 1$
	of the final driving amplitude.}    
\label{F_6}
\end{figure}

A variation of this theme is shown in Fig.~\ref{F_6}, which depicts the
corresponding results for $R_{\max}/(\hbar\omega) = 1$: Although the final
amplitude is four times smaller here than that employed in the preceding
Fig.~\ref{F_5}, so that the target state resembles the initial state more
closely, the decrease of the target-state population now becomes notable for
$z \approx 20$ already. The reason for this unexpected behavior can be spotted 
in Fig.~\ref{F_1}: With $R_{\max}/(\hbar\omega) = 1$ the final amplitude falls 
into the regime where ${\rm Var}_1(\hat{H}_{\rm drive}(t)/R)$ is large, 
signaling a relatively strong sensitivity of the Floquet states to the driving 
amplitude; this is felt by the state if it spends too much time in this regime.

\begin{figure}[t]
\centering
\includegraphics[width=0.9\linewidth]{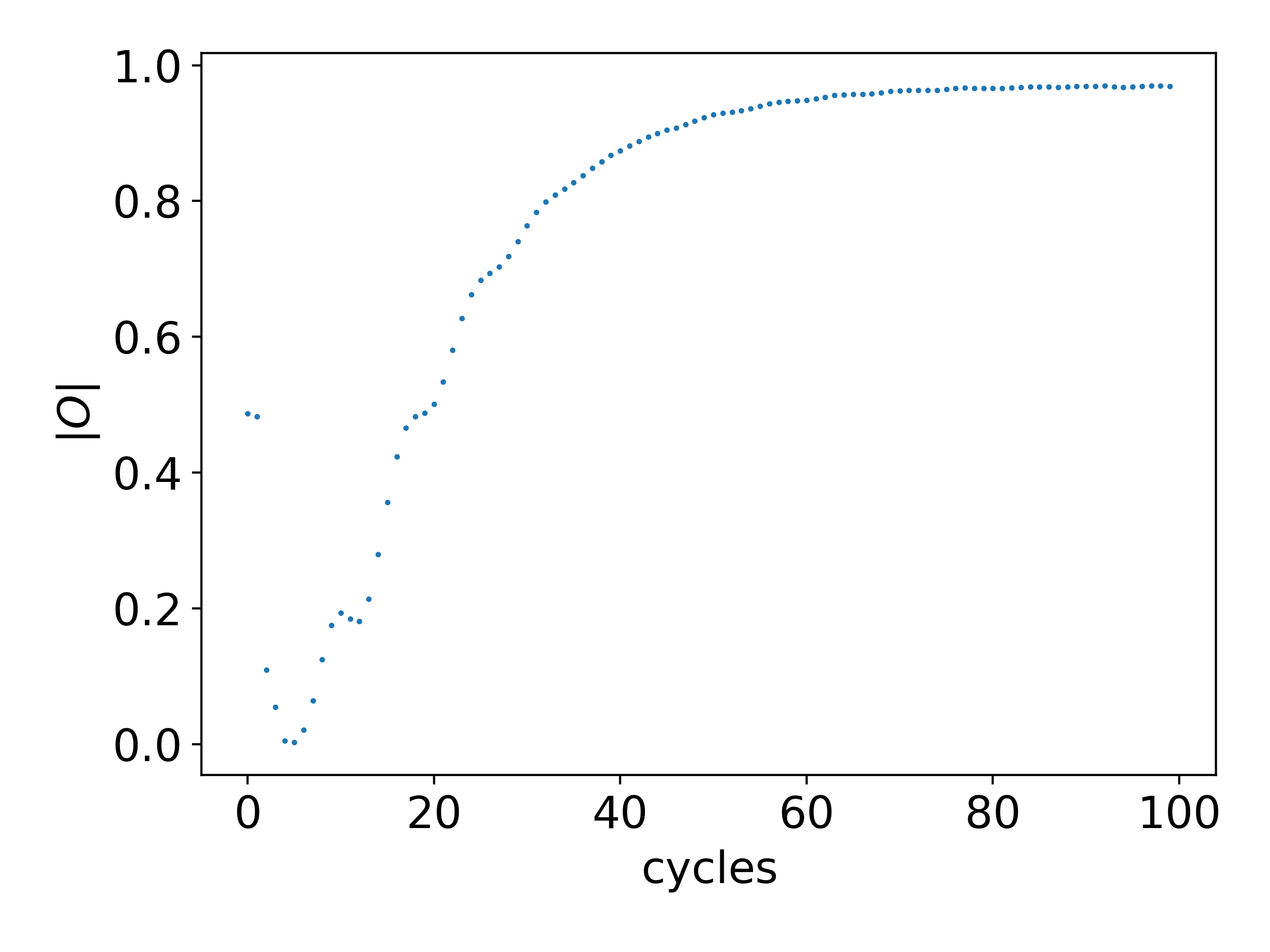}
\caption{As Fig.~\ref{F_5} but for higher value $R_{\max}/(\hbar\omega) = 7$
	of the final driving amplitude.}
\label{F_7}
\end{figure}

Finally, Fig.~\ref{F_7} shows data obtained for $R_{\max}/(\hbar\omega) = 7$,
beyond the second zero $j_{0,2}$ of ${\mathcal J}_0$. Here the overlap first
decreases with increasing length of the turn-on before it increases again. 
This behavior again may allow one to draw deductions concerning the 
instantaneous Floquet states involved: Now the initial state already exhibits 
a certain likeness to the target Floquet state, as is reflected by 
Fig.~\ref{F_3}; this likeness is advantageous in case of an almost sudden
turn-on. However, the time-evolving state is forced to undergo a substantial 
structural change for $j_{0,1} < R/(\hbar\omega) < j_{0,2}$, where the sign of 
the effective hopping matrix element~(\ref{eq:JEF}) is inverted, requiring 
a sufficiently moderate growth rate of the driving amplitude.

\section{Conclusions}
\label{S_7}

The iterative algorithm proposed in this work for calculating Floquet states 
of periodically time-dependent quantum systems does not yield the systems' 
full spectrum, but provides selected individual states and their quasienergy 
eigenvalues. Nonetheless, its attractive feature stems from the fact that it 
requires neither dia\-gonalization of the quasienergy operator in the 
extended Hilbert space~\cite{MurilloEtAl15} nor computation of the 
one-cycle evolution operator~\cite{LaptyevaEtAl16}. Thus, it provides a 
powerful tool for scrutinizing many-body Floquet systems that are not 
accessible to these older standard techniques due to the high dimensions of 
their Hilbert spaces.

Since a Floquet system does not possess a proper ground state, and since it 
is neither possible nor even desirable to compute all Floquet states of a 
truly high-dimensional system, one necessarily faces the question of how to 
select suitable, hopefully typical Floquet states for inspection. Here we 
have suggested to ``follow'' a Floquet state in parameter space by slightly 
modifying the quasienergy operator, seeding the iterative algorithm with the 
old Floquet state, and letting it relax to the new one; this strategy amounts 
to following according to maximum likeness of the Floquet states. 

Our explorative application of this strategy to a Bose-Hubbard chain subjected 
to a high-frequency drive was facilitated by the observation that here the
quasienergy which emerges from the system's ground-state energy undergoes only
tiny avoided crossings, unresolved on the scale of Fig.~\ref{F_2} when regarded
as function of the driving amplitude. In such favorable cases one encounters 
pseudoadiabatic Floquet state following in response to a driving amplitude 
that changes smoothly but still so fast that all these tiny avoided crossings
are traversed diabatically. The attempt to reach a formal adiabatic limit,
as corresponding to a change of the driving amplitude that proceeds
``infinitely slowly'', would be fruitless~\cite{HoneEtAl97} because countless 
Landau-Zener-like transitions would then divert the evolving state into 
undesired channels.

It should also be pointed out that the comparatively simple dynamics 
encountered in the present study within the high-frequency regime will give 
way to more complicated dynamics at lower frequencies, when resonances make 
themselves felt~\cite{PolettiKollath11}. With the tools provided here, which
do not rely on any high-frequency approximation, the Floquet analysis also of
such resonant dynamics can now be performed routinely.

From an experimental viewpoint, the paradigmatic results depicted in 
Figs.~\ref{F_5} -- \ref{F_7} suggest what may be termed ``turn-on  
spectroscopy'' with cold atoms in optical lattices: Prepare a superfluid 
or a Mott insulator state, then start shaking the lattice according to a 
deliberately designed turn-on protocol, and perform time-of-flight imaging 
after the shaking has become strictly periodic in time. Results obtained 
in this manner will depend on details of the respective turn-on protocol,
and model calculations of the type discussed in this paper may be of 
profound help for relating the different signatures observed for different 
protocols to the underlying quantum nonequilibrium many-body dynamics.

\begin{acknowledgments}
This work has been supported by the Deutsche For\-schungsgemeinschaft (DFG,
German Research Foundation) through Project No.~397122187. We thank the 
members of the Research Unit FOR~2692 for many stimulating discussions. 
\end{acknowledgments}

\end{document}